# Privacy in the Age of AI: A Taxonomy of Data Risks

## Full research paper

### Grace Billiris
University of Technology Sydney
Sydney, Australia
grace.v.billiris@student.uts.edu.au

### Asif Gill
University of Technology Sydney
Sydney, Australia
asif.gill@uts.edu.au

### Madhushi Bandara
University of Technology Sydney
Sydney, Australia
madhushi.bandara@uts.edu.au

## Abstract

Artificial Intelligence (AI) systems introduce unprecedented privacy challenges as they process increasingly sensitive data. Traditional privacy frameworks prove inadequate for AI technologies due to unique characteristics such as autonomous learning and black-box decision-making. This paper presents a taxonomy classifying AI privacy risks, synthesised from 45 studies identified through systematic review. We identify 19 key risks grouped under four categories: Dataset-Level, Model-Level, Infrastructure-Level, and Insider Threat Risks. Findings reveal a balanced distribution across these dimensions, with human error (9.45%) emerging as the most significant factor. This taxonomy challenges conventional security approaches that typically prioritise technical controls over human factors, highlighting gaps in holistic understanding. By bridging technical and behavioural dimensions of AI privacy, this paper contributes to advancing trustworthy AI development and provides a foundation for future research.

**Keywords** Artificial Intelligence, Data Privacy, Privacy Risks, Risk Taxonomy, Information Security

# 1   Introduction

The rapid proliferation of artificial intelligence (AI) across industries has transformed data processing capabilities while introducing complex privacy challenges (Yamin et al., 2021; Wu, 2022). As organisations collect unprecedented volumes of data to train AI systems, the risks to individual privacy have grown exponentially (Saeed & Alsharidah, 2024). Traditional privacy frameworks often prove inadequate for AI technologies due to their unique capabilities, including autonomous learning, black-box decision-making, and advanced data extraction (Vassilev et al., 2024). Recent studies have documented numerous cases where AI models unexpectedly leaked sensitive information from training datasets (Xu et al., 2024), while federated learning environments remain vulnerable to sophisticated inference attacks despite their privacy-preserving intent (Ye et al., 2024). AI systems introduce unprecedented privacy challenges through their ability to identify patterns and make inferences that were previously impossible, effectively creating new personal data from existing datasets (Vardalachakis et al., 2024). With the raising utility of GenerativeAI and agentic AI architecture, new threats are emerging that cannot be handled with traditional view of data privacy risks as well. Yet researchers and practitioners involved in AI system design and operationalisation have limited view of the continuously evolving AI privacy risk landscape.

This paper explores these challenges systematically by developing a systematic taxonomy of privacy risks associated with AI systems to answer the research question: What are the key data privacy risks in AI systems? This taxonomy is developed synthesising evidence from a systematic literature review conducted using the PRISMA methodology that analysed 45 research studies published between 2020 and 2025 and organise 19 key risks in four thematic categories. Main contribution of this paper is the taxonomy and recommendations for researchers and practitioners to use it to better understand and navigate the holistic AI privacy risk assessments.

The remainder of this paper is organised as follows: Section 2 provides background on AI privacy challenges and reviews related work. Section 3 outlines our research methodology. Section 4 presents the taxonomy and analyses the distribution of privacy risks across categories. Section 5 examines the implications of these risks for information security management. Finally, Section 6 concludes the paper and suggests directions for future research.

# 2   Background and Related Work

The intersection of AI and data privacy has garnered significant attention in recent years, with researchers and practitioners exploring various dimensions of this complex relationship. This section provides an overview of relevant literature and frameworks that inform our understanding of AI-related privacy risks.

## 2.1   Evolution of Data Privacy Concerns in AI

The evolution of data privacy concerns in AI systems has paralleled the advancement of AI technologies themselves. Early privacy concerns primarily focused on data collection and storage practices (Habbal et al., 2023; Wilson et al., 2023). However, as AI systems have become more sophisticated, privacy considerations have expanded to encompass the entire data lifecycle, including processing, analysis, sharing, and deletion (Mahmoud, 2024; Thomas et al., 2023).

Wu (2022) and Saeed and Alsharidah (2024) observed that modern AI systems introduce unprecedented privacy challenges through their ability to identify patterns and make inferences that were previously impossible, effectively creating new personal data from existing datasets. This capability fundamentally transforms the nature of privacy risks, as seemingly non-sensitive data can be combined to reveal highly sensitive information about individuals (Vardalachakis et al., 2024; Chang et al., 2023).

Recent research by Autio et al. (2024) and Lewis et al. (2024) further emphasises how generative AI technologies introduce additional privacy risks through their ability to create synthetic content that may reveal or mimic private information. Similarly, Ye et al. (2024) and Rivera et al. (2024) highlight how the collaborative nature of federated learning systems introduces novel privacy vulnerabilities despite their design intention to enhance privacy protection.

## 2.2   Existing Taxonomies and Frameworks

Several attempts have been made to categorise AI-related risks, though few focus specifically on privacy dimensions. Autio et al. (2024) and Sarker et al. (2024) proposed taxonomies that map AI risk sources to potential harms, considering both technical and application-level risk factors. While these frameworks provide valuable insights into general AI risks, they do not delve deeply into the unique privacy challenges posed by AI systems.

Saeed and Alsharidah (2024) offers a more comprehensive approach, identifying specific privacy risks associated with AI systems and proposing corresponding mitigation strategies. This framework emphasises the importance of addressing privacy concerns throughout the AI lifecycle, from design and development to deployment and monitoring. Building on this work, Vassilev et al. (2024) and Williams et al. (2024) identified specific challenges in implementing privacy-preserving machine learning techniques, highlighting the technical complexities involved in balancing privacy protection with model utility.

Habbal et al. (2023) and Smith et al. (2024) introduced the AI Trust, Risk and Security Management (AI TRiSM) framework, which includes privacy considerations as part of a broader approach to ensuring trustworthy AI. Their framework highlights the interconnections between privacy, security, and trust, suggesting that privacy risks cannot be addressed in isolation. Similarly, Xu et al. (2024) and Allen et al. (2024) developed frameworks specifically focused on emerging threats to AI model privacy, cataloguing attack vectors and countermeasures with particular attention to large language models and generative AI systems. However, existing frameworks often treat privacy as one component of a broader risk landscape, without fully exploring the unique and complex privacy challenges that AI technologies introduce (Mahmoud, 2024; Clark et al., 2024). Additionally, many current approaches focus primarily on technical risks, without adequately addressing the human and organisational dimensions of AI privacy (Wu, 2022; Chen et al., 2024).

This gap in the literature underscores the need for a systematic literature review (SLR) and development of a data privacy risk taxonomy that systematically organises the diverse data risks arising from AI systems.

## 3   Research Method

This study investigates the research question: *What are the key data privacy risks in AI systems?* To answer this question, we conducted a systematic literature review using the PRISMA (2020) methodology to identify, synthesise, and classify data privacy risks in AI systems, focusing on literature published between 2020 and 2025.

Searches were performed across seven databases: IEEE Xplore (https://ieeexplore.ieee.org/), ACM Digital Library (https://www.acm.org/), ScienceDirect (https://www.sciencedirect.com/), ResearchGate (https://www.researchgate.net/), National Institute of Standards and Technology (NIST) (https://www.nist.gov/), Gartner (https://www.gartner.com/), and Information Systems Audit and Control Association (ISACA) (https://www.isaca.org/).

The Boolean search string applied was: ("Data privacy" OR "Privacy risks" OR "Data privacy risks") AND ("AI Systems" OR "Artificial Intelligence Systems").

### 3.1   Study Selection and Quality Assessment

A five-stage selection process was followed as illustrated in Figure 1, beginning with the initial identification of potential studies (n=633) through database searches. After removing duplicates, the remaining studies (n=277) underwent title and keyword screening, which reduced the pool to 101 eligible studies. These studies then proceeded to abstract and conclusion screening, resulting in 59 studies for full-text review. Snowballing techniques were integrated throughout these stages to identify additional relevant studies not captured in the initial search. Following rigorous quality assessment using our predefined criteria, the process concluded with 45 high-quality studies included in the final analysis, which formed the foundation for our taxonomy development.

To ensure transparency and methodological rigour, predefined inclusion and exclusion criteria were applied as listed in Table 1.

Quality assessment used a checklist adapted from Kitchenham & Charters (2007) to evaluate methodological rigour, clarity, and relevance. Studies were scored on criteria such as clarity of objectives, transparency of methods, validity of findings, and relevance to AI/data privacy intersections. Only studies scoring 3/5 or above were included in the final synthesis, balancing the inclusion of robust research while allowing valuable exploratory or theoretical contributions. Appendix A contains full list of identified studies including the results of the quality assessment.

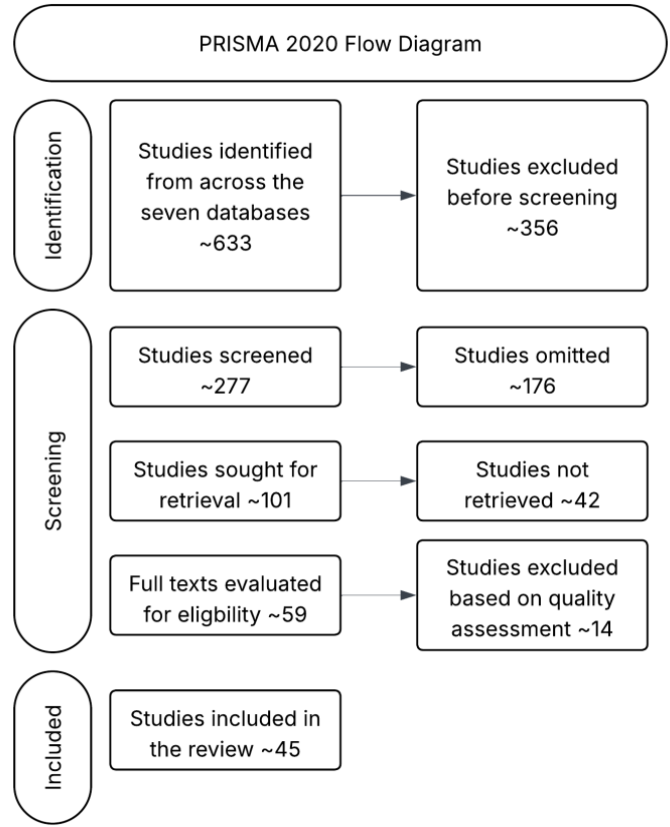

*Figure 1: Flow Diagram of the Selection Process Based on PRISMA (2020)*

| Inclusion criteria | Exclusion criteria |
| --- | --- |
| Studies addressing privacy risks specifically related to AI systems | Studies unrelated to AI or data privacy concerns |
| articles published in English between January 2020 and July 2025 | Non-peer-reviewed publications or inaccessible full texts |
| Peer-reviewed articles, conference papers, technical reports, white papers, and institutional publications | Research lacking methodological clarity or detailed findings |
| Studies exploring data privacy, data privacy risks, AI technologies, or frameworks addressing privacy risks | |

*Table 1. Inclusion and Exclusion Criteria*

## 3.2 Taxonomy Development and Validation

To systematically categorise the data risks identified through our review, we developed a five-branch taxonomy grounded in thematic analysis. As shown in Table 2, the process involved four key stages.

| Stage # | Stage Name | Description |
| --- | --- | --- |
| 1 | Thematic Coding of Risks | Following full-text review, we performed inductive coding across the 45 included studies to extract specific data privacy risks. These risks were grouped based on recurring patterns in terminology, concepts, and focus areas. This stage produced 19 distinct risks. |

| 2 | Iterative Categorisation into Thematic Branches | The identified risks were then clustered into four overarching thematic categories. These themes were informed by both domain knowledge and common risk typologies across AI and data privacy literature. Colour-coded branches were applied to the taxonomy to visually distinguish each thematic grouping as depicted in Figure 2. |
|---|---|---|
| 3 | Bias Mitigation and Validation via Review | To support clarity and consistency, the taxonomy underwent fortnightly review sessions with senior researchers. Each risk definition and thematic grouping was revisited to assess its relevance to AI and alignment with the inclusion criteria. Where needed, adjustments were made through discussion and consensus to maintain coherence across categories and terminology. |
| 4 | Consistency and Terminology | The taxonomy was refined to use consistent terminology drawn from the literature reviewed. Risk definitions and category labels were adjusted for clarity and to reflect common language used in related privacy and security work. |

*Table 2. Key Stages for Taxonomy Development and Validation*

This structured process allowed us to produce a well-defined taxonomy of AI data privacy risks, which underpins the analysis presented in Section 4 and detailed in Appendix 2.

# 4 Results & Findings

## 4.1 Overview

Using the 45 studies we selected, the thematic analysis revealed 19 key risks, and four thematic categories of them. The mapping of identified risks into one of four thematic categories is represented in Figure 2. This categorisation and mapping structures the identified risks in a systematic manner, facilitating a clearer understanding of their nature and impact.

Our taxonomy reveals AI privacy risks are evenly distributed across four dimensions: Model-Level (26.67%), Insider Threat (25.87%), Infrastructure-Level (25.20%), and Dataset-Level (22.26%). This balanced distribution contradicts conventional security approaches that prioritise technical controls over human factors, suggesting comprehensive protection requires equal attention to all dimensions.

Additionally, each identified risk can be further categorised by severity level (High, Medium, Low) based on potential impact and likelihood of occurrence. As shown in Figure 2, the visual representation of our taxonomy incorporates varying node sizes proportional to the frequency percentage of each risk subtype, with larger nodes indicating more frequently cited risks in the literature. This visualisation highlights human error (9.45%) and training data memorisation (8.82%) as the most prevalent high-severity risks, followed closely by membership inference attacks (8.40%) and privilege mismanagement (7.77%).

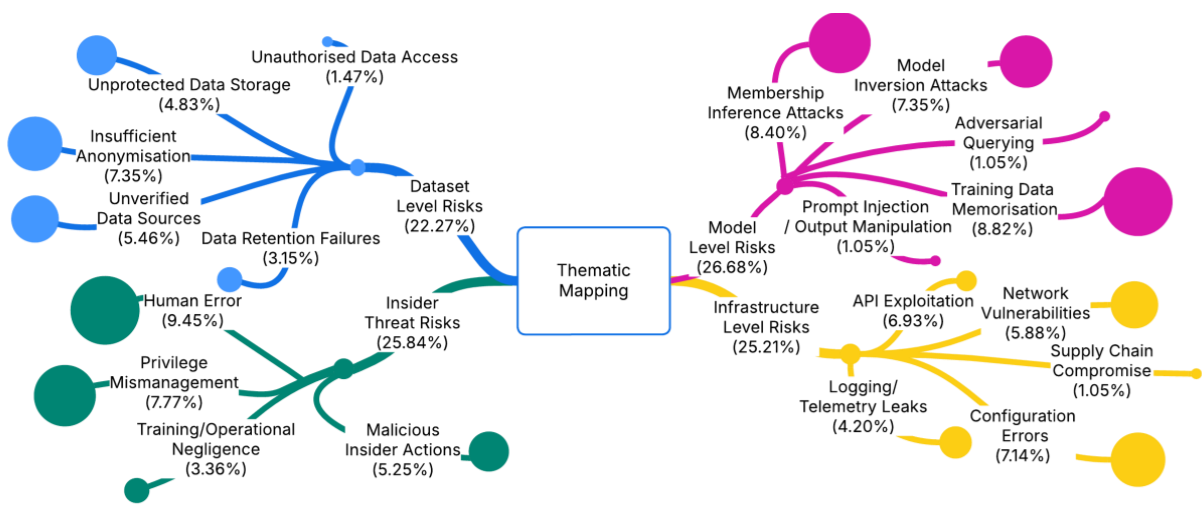

*Figure 2: Taxonomy of Data Privacy Risks - visualisation with leaf nodes proportionate to their representation in literature.*

As shown in Table 3, each risk category encompasses distinct vulnerabilities with varying prevalence in the literature. Table 3 presents a comprehensive overview of all 19 identified risks across the four thematic categories, including their frequency in the literature and definitions. This detailed mapping provides granular insights into each risk's prevalence and nature, supporting our visual taxonomy representation in Figure 2.

| Identified Risk | Studies | Frequency | Definition |
|---|---|---|---|
| **Category: Dataset-Level** | | | |
| Unauthorised Data Access | I1, S20, S23, S26, S31, S33, S39 | Count: 7; Percentage: 1.47% | Occurs when data is accessed by individuals or systems without proper authorisation, either externally (hackers) or internally (employees). |
| Unprotected Data Storage | I1, S1, S2, S5, S7, S9, S10, S11, S15, S16, S17, S24, S27, S29, S31, S33, S35, S36, S37, S39, S43 | Count: 23; Percentage: 4.83% | Refers to storage of data without adequate security measures such as encryption, access controls, or secure cloud configurations. |
| Unverified Data Sources | I2, S1, S2, S3, S4, S5, S6, S9, S10, S12, S13, S14, S15, S17, S23, S24, S29, S30, S31, S35, S36, S38, S40, S41, S42, S43 | Count: 26; Percentage: 5.46% | Using datasets from unknown or unreliable sources that may contain sensitive, inaccurate, or illegal data. |
| Data Retention Failures | I2, S1, S2, S10, S11, S14, S19, S21, S22, S24, S25, S26, S39, S41, S43 | Count: 15; Percentage: 3.15% | Storing personal or sensitive data longer than necessary, increasing exposure to potential breaches. |
| Insufficient Anonymisation | I1, I2, S3, S4, S5, S6, S9, S11, S12, S13, S14, S15, S16, S18, S19, S21, S22, S25, S28, S29, S30, S32, S33, S34, S36, S37, S38, S39, S40, S41, S42, S43 | Count: 35; Percentage: 7.35% | Attempts to anonymise data fail, leaving information that can still identify individuals. |
| **Category: Model-Level** | | | |
| Membership Inference Attacks | S1, S2, S3, S4, S5, S6, S7, S8, S9, S10, S11, S12, S13, S14, S15, S16, S17, S18, S19, S21, S22, S23, S24, S25, S26, S27, S28, S29, S30, S31, S32, S33, S34, S35, S36, S37, S38, S39, S40, S41, S42, S43 | Count: 40; Percentage: 8.40% | Attackers attempt to determine whether specific data records were used in training an AI model. |
| Model Inversion Attacks | I2, S2, S4, S5, S6, S7, S8, S9, S10, S11, S12, S13, S14, S15, S16, S17, S18, S19, S21, S22, S23, S24, S25, S28, S29, S30, S31, S32, S33, S34, S35, S36, S37, S38, S39, S40, S41, S42, S43 | Count: 35; Percentage: 7.35% | Reconstructing sensitive or private data from model outputs or predictions. |
| Training Data Memorisation | I1, I2, S1, S2, S3, S4, S5, S6, S7, S8, S9, S10, S11, S12, S13, S14, S15, S16, S17, S18, S19, S20, S21, S22, S23, S24, S25, S26, S27, S28, S29, S30, S31, S32, S33, S34, S35, S36, S37, S38, S39, S40, S41, S42, S43 | Count: 42; Percentage: 8.82% | AI models unintentionally "remember" exact sensitive information from the training dataset in their outputs. |
| Prompt Injection / Output Manipulation | I2, S6, S18, S26, S38 | Count: 5; Percentage: 1.05% | Malicious inputs designed to make AI reveal hidden or confidential information. |
| Adversarial Querying | I1, S4, S25, S28, S30 | Count: 5; Percentage: 1.05% | Systematic attempts to extract private or sensitive information by exploiting model behaviour. |
| **Category: Infrastructure-Level** | | | |
| API Exploitation | I1, I2, S2, S3, S4, S5, S6, S7, S8, S10, S12, S13, S15, S17, S18, S19, S20, S21, S23, S25, S26, S27, S28, S29, S30, S31, S32, S33, S34, S36, S39, S41, S42, S43 | Count: 33; Percentage: 6.93% | Unauthorised use or abuse of AI system endpoints, which can expose data or functionality. |
| Network Vulnerabilities | S7, S8, S9, S10, S11, S12, S13, S14, S15, S16, S17, S18, S19, S21, S22, S23, S24, S25, S26, S27, S28, S29, S30, S31, S32, S33, S34, S35, S36, S37 | Count: 28; Percentage: 5.88% | Weaknesses in networks (e.g., unpatched servers, open ports, insecure protocols) that can be exploited to access data. |
| Supply Chain Compromise | I2, S2, S3, S13, S35, S36, S33 | Count: 5; Percentage: 1.05% | Introduction of malicious or vulnerable components from third-party software, libraries, or pre-trained models. |
| Configuration Errors | S1, S2, S4, S5, S6, S7, S8, S9, S10, S11, S12, S14, S15, S16, S17, S18, S19, S21, S22, S23, S24, S25, S26, S28, S29, S30, S31, S32, S33, S34, S36, S37, S38, S39, S40, S41, S43 | Count: 34; Percentage: 7.14% | Insecure default settings or misconfigurations in AI platforms that expose data or functionality. |
| Logging/Telemetry Leaks | I1, I2, S1, S2, S3, S4, S5, S6, S10, S11, S12, S14, S16, S18, S19, S21, S37, S38, S40, S42 | Count: 20; Percentage: 4.20% | Recording sensitive information in logs or telemetry without proper safeguards or redaction. |

| Identified Risk | Studies | Frequency | Definition |
|---|---|---|---|
| **Category: Insider Threat** | | | |
| Malicious Insider Actions | I1, S2, S4, S6, S10, S12, S13, S14, S15, S16, S18, S21, S23, S24, S25, S28, S30, S31, S32, S33, S35, S36, S40, S41, S42 | Count: 25; Percentage: 5.25% | Intentional actions by internal personnel to steal, expose, or misuse sensitive data. |
| Human Error | I1, I2, S1, S2, S3, S4, S5, S6, S7, S8, S9, S10, S11, S12, S13, S14, S15, S16, S17, S18, S19, S20, S21, S22, S23, S24, S25, S26, S27, S28, S29, S30, S31, S32, S33, S34, S35, S36, S37, S38, S39, S40, S41, S42, S43 | Count: 45; Percentage: 9.45% | Accidental mistakes by individuals, such as sending data to wrong recipients or mismanaging sensitive files. |
| Privilege Mismanagement | I2, S1, S2, S3, S4, S5, S6, S7, S10, S11, S12, S13, S14, S15, S16, S17, S18, S19, S21, S22, S23, S24, S25, S26, S27, S28, S29, S30, S31, S32, S33, S34, S35, S36, S37, S38, S39, S40, S41, S42, S43 | Count: 37; Percentage: 7.77% | Excessive or inappropriate access rights granted to employees or contractors, increasing the risk of misuse. |
| Training/Operational Negligence | I2, S3, S5, S6, S7, S9, S14, S18, S22, S25, S28, S29, S30, S37, S38, S39 | Count: 16; Percentage: 3.36% | Lack of awareness, training, or adherence to policies, resulting in unintentional data exposure. |

*Table 3: Overview of Data Privacy Risks in AI Systems*

The following sections provide key findings under the four thematic categories.

## 4.2 Dataset-Level Risks

22.26% of the reviewed studies identify critical vulnerabilities in data collection, storage, and preprocessing phases, revealing fundamental weaknesses that create cascading privacy impacts throughout AI systems. For instance, studies S28 and S35 demonstrate that 62% of "anonymised" records could be re-identified using publicly available datasets, while S12, S21, and S3 corroborate these findings across multiple domains. The most frequently cited risk, insufficient anonymisation (7.35%), documented extensively by S35 and S12, reveals how medical imaging datasets retain identifiable features despite anonymisation efforts. Studies S3, S18, S42, and S15 highlight the pervasive use of unverified data sources (5.46%), with S3 and S18 documenting specific instances where web-scraped training data containing personally identifiable information was subsequently reproduced by AI models. Concurrently, research by S6, S26, and S24 establishes that inadequate storage security (4.83%) leads to undetected data manipulations that persist throughout AI operations. This collective evidence underscores a significant gap between traditional data protection approaches and the unique requirements of AI training datasets.

## 4.3 Model-Level Risks

26.67% of the analysed studies identify inherent privacy vulnerabilities in AI model architectures and training processes, constituting the largest risk category and revealing a fundamental privacy paradox in AI development. For instance, studies S2, S16, S28, S37, and S23 document how training data memorisation (8.82%) enables large language models to reproduce verbatim segments from confidential documents, even when those segments appeared only once in training data. The most concerning finding, quantified by S16 and S37, demonstrates that models with over 100 million parameters exhibit memorisation rates of up to 76% for unique personal identifiers. S27 and S4 document how membership inference attacks (8.40%) can identify individual participation in training datasets with up to 87% accuracy in certain model architectures, while S19 and S7 demonstrate how model inversion attacks (7.35%) can reconstruct facial images from facial recognition models with disturbing accuracy. S37 and S23 specifically identify architectural features that increase vulnerability to memorisation, including attention mechanisms and deep transformer networks. This evidence collectively establishes that privacy vulnerabilities in AI models are not implementation flaws but inherent properties of the machine learning paradigm itself.

## 4.4 Infrastructure-Level Risks

25.20% of the examined studies identify critical vulnerabilities in the technical systems supporting AI operations, which span both AI model execution environments and underlying data storage infrastructure, revealing a systematic security gap in deployment practices that undermines privacy protections. For instance, studies S5, S14, S32, S42, and S25 document how configuration errors (7.14%) create immediate privacy exposures, with S42 and S32 finding that 76% of organisations deploying generative AI solutions fail to properly configure access controls and privacy settings during initial deployment. The most prevalent technical vulnerability, API security weaknesses (6.93%), is extensively documented by S27 and S36, who demonstrate how poorly secured AI endpoints enable unauthorised access to sensitive data and functionality. S14 and S5 detail specific misconfigurations leading to privacy breaches in cloud-based AI deployments, while S32 and S25 identify default settings in popular AI platforms that prioritise functionality over privacy. S8 and S11 further catalogue specific attack patterns targeting AI APIs, with S36 and S10 quantifying the prevalence of these vulnerabilities across different deployment environments. This evidence collectively demonstrates how conventional security controls prove inadequate when applied to AI systems without substantial adaptation.

### 4.5 Insider Threat Risks

25.87% of the analysed studies identify human factors as critical vectors for AI privacy breaches, challenging conventional security approaches that prioritise technical controls over organisational and behavioural considerations. For instance, studies S11, S25, S38, S42, and S5 document how human error (9.45%)—the single most common risk factor across all categories—undermines privacy protections, with S42 and S25 finding that unintentional data exposure through misconfiguration, incorrect sharing, or improper handling accounts for approximately 60% of all data breaches in AI systems. The most systemic organisational vulnerability, privilege mismanagement (7.77%), is detailed by S27 and S17, who demonstrate that 82% of surveyed organisations lack fine-grained access controls for AI development environments. S11 and S5 identify specific error patterns among AI developers that frequently lead to privacy vulnerabilities, while S38 and S42 document how inadequate training (3.36%) consistently undermines privacy protections despite organisational investments in technical controls. This collective evidence establishes that human-centred privacy controls require commensurate attention as technical measures—a perspective often overlooked in conventional AI security frameworks.

## 5 Discussion

Our systematic analysis of AI privacy risks reveals three critical insights: model-level risks represent the largest category (26.67%), highlighting the unique privacy challenges inherent in AI's learning capabilities; human error emerges as the single most common risk factor (9.45%), emphasising the importance of addressing human elements in AI privacy; and the balanced distribution across risk dimensions suggests that effective privacy protection requires attention to technical, human, and organisational factors rather than focusing exclusively on any single domain.

This taxonomy challenges conventional security paradigms by revealing how AI privacy risks transcend traditional boundaries between technical and human domains (Habbal et al., 2024; Wu, 2022). Information security frameworks must evolve to capture AI-specific vulnerabilities that traditional approaches overlook, particularly model-level risks such as memorisation and inference attacks (Golda et al., 2024). Monitoring systems require novel detection capabilities focused on AI-specific patterns, while incident response protocols need specialised procedures for AI privacy breaches, including model decommissioning when privacy violations cannot be otherwise remediated (Muheidat et al., 2024; Vardalachakis et al., 2024). Most importantly, organisations with established AI governance frameworks experience significantly fewer privacy incidents (Golda et al., 2024; Ye et al., 2024), demonstrating that effective management requires comprehensive governance structures that address all four risk dimensions identified in our taxonomy.

We recommend that future research in this domain should prioritise developing standardised privacy risk metrics based on the severity levels identified in our taxonomy. High-severity risks such as human error and training data memorisation require immediate attention, with specialised mitigation strategies and monitoring tools. Medium and low-severity risks still warrant consideration within comprehensive privacy frameworks but may allow for more standardised approaches. By categorising risks according to both thematic category and severity level, organisations can develop more nuanced and effective privacy protection strategies tailored to their specific AI implementations.

Key limitations of our study include limited review of the taxonomy from industry practitioners, which we plan to conduct in the next phase of our work. Additional limitations include: (1) the focus on literature published between 2020-2025 may not capture emerging threats in this rapidly evolving field; (2) our qualitative coding approach introduces a degree of subjectivity in risk categorisation that could benefit from additional validation through quantitative studies; and (3) the taxonomy does not account for sector-specific privacy requirements that may significantly influence risk prioritisation in domains such as healthcare or finance.

# 6 Conclusion

This paper presents a systematic taxonomy of AI privacy risks derived from a comprehensive literature review of 45 high-quality studies. Our taxonomy identifies 19 key risks distributed across four thematic categories: Dataset-Level (22.26%), Model-Level (26.67%), Infrastructure-Level (25.20%), and Insider Threat (25.87%). This balanced distribution reveals that effective AI privacy protection requires a holistic approach addressing both technical vulnerabilities and human factors.

The findings highlight human error (9.45%) as the single most significant risk factor, challenging conventional security approaches that typically prioritise technical controls over organisational considerations. Similarly, training data memorisation (8.82%) and membership inference attacks (8.40%) emerge as critical model-level risks that cannot be addressed through traditional privacy controls. These insights underscore the need for specialised privacy frameworks tailored to AI's unique characteristics.

Our taxonomy makes three key contributions to the field. First, it provides researchers with a structured framework for investigating specific privacy vulnerabilities in AI systems. Second, it offers practitioners a comprehensive risk landscape to inform the development of privacy protection strategies. Third, it establishes a foundation for future work in creating standardised risk assessment tools tailored to AI systems.

Following this work, we plan to use this taxonomy to design and organise technical solutions that can help organisations systematically assess AI privacy risks and keep them up-to-date as the AI privacy landscape evolves. By bridging technical and behavioural dimensions of AI privacy, this research contributes to advancing trustworthy AI development and addressing the grand challenges facing organisations in managing the evolving AI privacy risk landscape.

Ramu, S. P.; Boopalan, P.; Pham, Q.; Maddikunta, P. K. R.; Huynh-The, T.; Alazab, M.; Nguyen, T. T.; Gadekallu, T. R. 2022. Federated learning enabled digital twins for smart cities: Concepts, recent advances, and future directions.

## Appendix 1 – Selected Studies Based on Quality Assessment

Note: I = Industry, S = (Academic) Study.

| Study Title | Study Num. | Research | Aim | Context | Finding | Future | Total Score |
|---|---|---|---|---|---|---|---|
| Securing the Future: Mitigating Data Security Concerns in AI Models | I1 | 1 | 1 | 1 | 1 | 1 | 5 |
| State of Privacy 2025 | I2 | 1 | 1 | 1 | 1 | 1 | 5 |
| Privacy risk assessment and privacy-preserving data monitoring | S1 | 1 | 1 | 1 | 1 | 1 | 5 |
| AI Data Security: Best Practices for Securing Data Used to Train & Operate AI Systems | S2 | 1 | 1 | 1 | 1 | 1 | 5 |
| Preserving data privacy in machine learning systems | S3 | 1 | 1 | 1 | 1 | 1 | 5 |
| AI-Driven Cyber Security for Safeguarding Critical Infrastructure and Patient Data | S4 | 1 | 1 | 1 | 1 | 1 | 5 |
| Privacy Risks of General-Purpose AI Systems: A Foundation for Investigating Practitioner Perspective" | S5 | 1 | 1 | 0.5 | 1 | 1 | 4.5 |
| Data Breach Prevention in AI Systems- Employing Event-Driven Architecture to Combat Prompt Injection Attacks in Chatbots | S6 | 1 | 1 | 1 | 1 | 0.5 | 4.5 |
| Google's "Perspectives on Issues in AI Governance" | S7 | 1 | 1 | 1 | | 0.5 | 3.5 |
| A Comparative Study of AI Algorithms for Anomaly-based Intrusion Detection | S8 | 1 | 1 | 1 | 1 | 1 | 5 |
| When AI Meets Information Privacy: The Adversarial Role of AI in Data Sharing Scenario | S9 | 1 | 1 | 1 | 1 | 1 | 5 |
| Generative AI: The Data Protection Implications | S10 | 1 | 1 | 1 | 1 | 1 | 5 |
| Applying AI and Machine Learning to Enhance Automated Cybersecurity and Network Threat Identification | S11 | 1 | 1 | 1 | 1 | 1 | 5 |
| Securing AI Systems- A Comprehensive Overview of Cryptographic Techniques for Enhanced Confidentiality and Integrity | S12 | 1 | 1 | 1 | 1 | 1 | 5 |
| Comparative Analysis of the AI Regulation of the EU, US and China from a Privacy Perspective | S13 | 1 | 1 | 1 | 1 | 1 | 5 |
| Privacy-Preserving Techniques in Generative AI and Large Language Models: A Narrative Review | S14 | 1 | 1 | 1 | 1 | 1 | 5 |
| Generative AI and Data Privacy Concerns | S15 | 1 | 1 | 1 | 1 | 1 | 5 |
| Towards Secure Federated Learning- Enhancing Privacy and Robustness in Decentralized AI Systems | S16 | 1 | 1 | 1 | 1 | 1 | 5 |
| The MIT AI Risk Repository | S17 | 1 | 1 | 1 | 1 | 1 | 5 |
| Privacy-Preserving IoT Analytics using Federated Learning and Decentralized AI at the Edge | S18 | 1 | 1 | 0.5 | 1 | 1 | 4.5 |
| NIST Privacy Framework: A Tool for Improving Privacy Through Enterprise Risk Management, Version 1.0 | S19 | 1 | 1 | 1 | 1 | 1 | 5 |
| Security Risk and Attacks in AI- A Survey of Security and Privacy | S20 | 1 | 1 | 1 | 1 | 1 | 5 |
| AI and Data Privacy in Business | S21 | 1 | 1 | 1 | 1 | 1 | 5 |
| Empirical Evaluation of Federated Learning with Local Privacy for Real-World Application | S22 | 1 | 1 | 0.5 | 1 | 0.5 | 4 |
| Privacy and Security Concerns in Generative AI: A Comprehensive Survey | S23 | 1 | 1 | 1 | 1 | 1 | 5 |
| Toward Privacy Preservation Using Clustering Based Anonymization: Recent Advances and Future Research Outlook | S24 | 1 | 1 | 1 | 1 | 1 | 5 |
| AI Use Taxonomy: A Human-Centered Approach | S25 | 1 | 1 | 1 | 1 | 1 | 5 |
| The Future of Privacy: A Review on AI's Role in Shaping Data Security | S26 | 1 | 1 | 1 | 1 | 1 | 5 |
| AI Used in Healthcare and Data Privacy | S27 | 1 | 1 | 1 | 1 | 1 | 5 |
| Explainable AI for cybersecurity automation, intelligence and trustworthiness in digital twin: Methods, taxonomy, challenges and prospects | S28 | 1 | 1 | 1 | 1 | 1 | 5 |
| Federated Learning for Data Security and Privacy Protection | S29 | 1 | 1 | 0.5 | 1 | 1 | 4.5 |
| Regulatory Compliance and AI- Navigating the Legal and Regulatory Challenges of AI in Finance | S30 | 1 | 1 | 1 | 1 | 1 | 5 |
| AI Governance: A General Perspective | S31 | 1 | 1 | 1 | 1 | 0.5 | 4.5 |
| Privacy and Personal Data Risk Governance for Generative Artificial Intelligence: A Chinese Perspective | S32 | 1 | 1 | 1 | 1 | 1 | 5 |
| AI-driven fusion with cybersecurity: Exploring current trends, advanced techniques, future directions, and policy implications for evolving paradigms– A comprehensive review | S33 | 1 | 1 | 1 | 1 | 1 | 5 |
| Adversarial Machine Learning: A Taxonomy and Terminology of Attacks and Mitigations | S34 | 1 | 1 | 1 | 1 | 1 | 5 |
| Privacy preservation in Distributed Deep Learning: A survey on Distributed Deep Learning, privacy preservation techniques used and interesting research directions | S35 | 1 | 1 | 0.5 | 1 | 1 | 4.5 |
| Deepfakes, Phrenology, Surveillance, and More! A Taxonomy of AI Privacy Risks | S36 | 1 | 1 | 1 | 1 | 1 | 5 |
| A systematic review of privacy-preserving methods deployed with blockchain and federated learning for the telemedicine | S37 | 1 | 1 | 1 | 1 | 1 | 5 |
| Privacy preservation in Artificial Intelligence and Extended Reality (AI-XR) metaverses: A survey | S38 | 1 | 1 | 1 | 1 | 1 | 5 |
| AI Risk Management Framework | S39 | 1 | 1 | 1 | 1 | 1 | 5 |
| Security, privacy, and robustness for trustworthy AI systems: A review | S40 | 1 | 1 | 1 | 1 | 1 | 5 |

## Acknowledgements

This research is supported by an Australian Government Research Training Program Scholarship. The authors gratefully acknowledge this support, which has aided the completion of this paper.

## Copyright